\documentclass[10pt]{article}

\usepackage[dviwin]{graphicx}
\usepackage{amsfonts}
\usepackage{amssymb}
\usepackage{mathrsfs}
\usepackage{amsmath}
\usepackage{indentfirst}
\usepackage{dsfont}
\usepackage{esint}
\usepackage{hyperref}
\hypersetup{
    colorlinks = true,
    linkcolor = [rgb]{0 .4 .8},
    citecolor=[rgb]{1 .2 .2}
}
\usepackage{array}
\usepackage[numbers,sort]{natbib}
\setcitestyle{square}

\newcommand{\bea}{\begin{equation}}
\newcommand{\eea}{\end{equation}}
\newcommand{\bear}{\begin{eqnarray}}
\newcommand{\eear}{\end{eqnarray}}
\newcommand{\bearr}{\begin{eqnarray*}}
\newcommand{\eearr}{\end{eqnarray*}}

\newcommand{\beal}{\begin{align}}
\newcommand{\eeal}{\end{align}}
\newcommand{\beall}{\begin{align*}}
\newcommand{\eeall}{\end{align*}}

\newcommand{\tr}{\mathrm{tr}\,}
\newcommand{\CP}{\mathds{C}\mathds{P}}
\newcommand{\CC}{\mathds{C}}

\newcommand{\dd}{\partial}

\newcommand{\comment}[1]{}

\usepackage{xcolor}
\usepackage{empheq}

\newcolumntype{L}[1]{>{\raggedright\let\newline\\\arraybackslash\hspace{0pt}}m{#1}}
\newcolumntype{C}[1]{>{\centering\let\newline\\\arraybackslash\hspace{0pt}}m{#1}}
\newcolumntype{R}[1]{>{\raggedleft\let\newline\\\arraybackslash\hspace{0pt}}m{#1}}

\makeatletter
\def\@seccntformat#1{\@ifundefined{#1@cntformat}%
{\csname the#1\endcsname\quad}
{\csname #1@cntformat\endcsname}
}
\def\section@cntformat{{\normalfont\large\thesection.}\quad}
\def\subsection@cntformat{\textsection\, \thesubsection.\quad}
\def\subsubsection@cntformat{\textsection\textsection\, \thesubsubsection.\quad}
\makeatother

\newsavebox\MBox

\newcommand*{\TitleFont}{%
      \usefont{\encodingdefault}{\rmdefault}{b}{n}%
      \fontsize{14}{20}%
      \selectfont}
 
\usepackage{sectsty}
 
\usepackage{bm}

\DeclareBoldMathCommand\boldlangle{\left\langle}
\DeclareBoldMathCommand\boldrangle{\right\rangle}

\sectionfont{\fontsize{12}{15}\selectfont}

\begin{document}
\vspace{-1cm}
\title{\vspace{-1.5cm}\TitleFont Complex structures and \\Zero-curvature equations for $\sigma$-models}
\author{Dmitri Bykov\footnote{Emails:
dmitri.bykov@aei.mpg.de, dbykov@mi.ras.ru}  \\ \\
{\small $\bullet$ Max-Planck-Institut f\"ur Gravitationsphysik, Albert-Einstein-Institut,} \\ {\small Am M\"uhlenberg 1, D-14476 Potsdam-Golm, Germany} \\ {\small $\bullet$ Steklov
Mathematical Institute of Russ. Acad. Sci.,}\\ {\small Gubkina str. 8, 119991 Moscow, Russia \;}}
\date{}

\maketitle
\vspace{-1cm}
\begin{center}
\line(1,0){370}
\end{center}
\vspace{-0.2cm}
\textbf{Abstract.} We construct zero-curvature representations for the equations of motion of a class of $\sigma$-models with complex homogeneous target spaces, not necessarily symmetric. We show that in the symmetric case the proposed flat connection is gauge-equivalent to the conventional one.
\vspace{-0.7cm}
\begin{center}
\line(1,0){370}
\end{center}

\section{The models}\label{lagrsec}

A $\sigma$-model is a field theory describing maps $X: \Sigma \to \mathcal{M}$ from a worldsheet $\Sigma$ to a target space $\mathcal{M}$. In this paper the worldsheet $\Sigma$ will be a two-dimensional Euclidean manifold, and the target-space $\mathcal{M}$ will be required to have the following properties\footnote{Generalizations to non-simple groups $G$ are possible, but will be considered elsewhere.}:
\begin{align} \nonumber
&\circ\quad \mathcal{M}\, \textrm{is a homogeneous space}\; \;G/H, \;\;G\; \textrm{semi-simple and compact} \\ \label{prop}
&\circ\quad \mathcal{M}\, \textrm{has an integrable}\; G\textrm{-invariant complex structure}\; \mathbb{I} \\
\nonumber
&\circ\quad \textrm{The Killing metric}\; \mathbb{G}\; \textrm{on}\; \mathcal{M}\; \textrm{is Hermitian w.r.t.}\; \mathbb{I} 
\end{align}
We will show that, for a target space $\mathcal{M}$ with these properties, one can define a \mbox{$\sigma$-model,} whose equations of motion may be rewritten as the flatness condition for a one-parameter family of connections $A_u, u\in \CC^\ast$. This is an extension to a broader class of target spaces of a property that is encountered in $\sigma$-models with symmetric target spaces \cite{Pohlmeyer, Forger}. In the latter case it is an important sign of integrability of the model: it may be used to find B\"acklund transformations \cite{Uhlenbeck, Devchand}, and it is a starting point for the construction of classical solutions of the models \cite{Hitchin}.

\vspace{0.3cm}
Complex simply-connected homogeneous manifolds $G/H$ with $G$ semi-simple were classified long ago \cite{Wang}. They are given by the following theorem: any such manifold $G/H$ corresponds to a subgroup $H$, whose semi-simple part coincides with the semi-simple part of the centralizer of a toric subgroup of $G$.

For the case of $G=SU(N)$, for example, invariant complex structures exist on those of the  manifolds
\bea\label{complhom}
\mathcal{M}_{n_1, \ldots, n_m | N}=\frac{SU(N)}{S(U(n_1)\times \ldots \times U(n_m))},\quad\quad m\geq 0, \;\;n_i> 0,\;\;\sum\limits_{i=1}^m \,n_i\leq N\,,
\eea
that are even-dimensional. If $\sum_{i=1}^m \,n_i= N$, the manifold in (\ref{complhom}) is a flag manifold (For a review of flag manifolds see \cite{Alekseevsky, Arvanito}). Otherwise, it is a toric bundle over a flag manifold. The fiber $U(1)^{2s}$ ($2s=N-\sum_{i=1}^m \,n_i$) is even-dimensional, since the flag manifold itself is even-dimensional.

\vspace{0.3cm}
If one relaxes the requirements of simple connectedness of $G/H$ and semi-simplicity of $G$ in the above theorem, there are more examples, such as $S^1\times S^3=U(2)$, which is even hyper-complex, i.e. it has a continuous family of invariant complex structures (such manifolds were considered and classified in \cite{Troost} and \cite{Joyce}). In this case, in place of the Killing metric one can take the metric $\widehat{\mathbb{G}}_{ij}:=\tr(T_iT_j)$, where $T_i$ are the generators of $\mathfrak{g}$ taken in a representation suitably chosen in order to make sure that $\widehat{\mathbb{G}}$ is non-degenerate and Hermitian with respect to $\mathbb{I}$.

\vspace{0.3cm}
The models, which will be of interest for us in the present paper, are defined by the following action:
\bea\label{action}
\mathcal{S}[\mathbb{G}, \mathbb{I}]:=\int_\Sigma\,d^2 z\,\|\dd X\|^2_\mathbb{G}+\int_\Sigma\,X^\ast \omega,
\eea
where $\omega$ is the K\"ahler form corresponding to the pair $(\mathbb{G}, \mathbb{I})$, defined as
\bea\label{kahform}
\omega=\mathbb{G}\circ \mathbb{I}\,.
\eea
In general, the Killing metric $\mathbb{G}$ is not K\"ahler, i.e. the K\"ahler form is not closed: $d\omega \neq 0$. This might be the case, even if the manifold $\mathcal{M}$ admits a K\"ahler metric -- it is in general different from $\mathbb{G}$. As an example of such phenomenon one can consider the flag manifold $SU(3)\over S(U(1)^3)$. The $\sigma$-model (\ref{action}) for the flag manifold was investigated in detail in \cite{Bykov1, Bykov2}. Other examples of models of the class (\ref{prop}) are provided by Hermitian symmetric spaces -- symmetric spaces with complex structure. These manifolds are K\"ahler, and the invariant metric is essentially unique (up to scale), thus leading to the closedness of $\omega$: $d\omega=0$. We will discuss this special case in Section \ref{symmspace}.

\section{Structure of the target space}

The Lie algebra $\mathfrak{g}$ of the Lie group $G$ may be decomposed as follows: $\mathfrak{g}=\mathfrak{h}\oplus \mathfrak{m}$, where $\mathfrak{m}$ is the orthogonal complement to $\mathfrak{h}$ with respect to the Killing metric $\mathbb{G}$. We will assume that the quotient space $G/H$ possesses an almost complex structure $\mathbb{I}$. We are not postulating that $\mathbb{I}$ be integrable -- this will rather follow from the requirement of the existence of a Lax connection. The almost complex structure acts on $\mathfrak{m}$ and may be diagonalized, its eigenvalues being $\pm i$. We denote the $\pm i$-eigenspaces by $\mathfrak{m}_{\pm}$:
\bea\label{liealgdec}
\mathfrak{g}=\mathfrak{h}\oplus \mathfrak{m}_+\oplus \mathfrak{m}_-,\quad\quad \mathbb{I}\circ \mathfrak{m}_\pm=\pm i\,\mathfrak{m}_\pm\,.
\eea
$G$-invariance of the almost complex structure implies that $[\mathfrak{h}, \mathfrak{m}_\pm]\subset \mathfrak{m}_\pm$. We introduce the current
\bea
J=g^{-1}dg=J_0+J_++J_-,\quad\quad J_0\in\mathfrak{h},\;\;\;J_\pm\in \mathfrak{m}_\pm\,.
\eea
It takes values in the Lie algebra $\mathfrak{g}$, therefore we have decomposed it according to the decomposition (\ref{liealgdec}) of the Lie algebra. In these terms the action (\ref{action}) may be rewritten as follows:
\bea
\mathcal{S}[\mathbb{G}, \mathbb{I}]:=\int_\Sigma\,d^2 z\,\;\boldlangle (J_+)_z,\,(J_-)_{\bar{z}}\boldrangle_{\mathbb{G}}\;.
\eea
Henceforth we will be using bracket notation for the scalar product of two elements $\alpha, \beta \in \mathfrak{g}$: $\boldlangle \alpha, \beta\boldrangle_{\mathbb{G}}:=\mathbb{G}(\alpha, \beta)$. The Noether current, constructed using the above action, will be denoted by $K$. It is derived by taking an infinitesimal $(z, \bar{z})$-dependent variation $g\to e^{\epsilon(z, \bar{z})}\circ g$ in the above action, which leads to
\bear\nonumber
\delta\mathcal{S}[\mathbb{G}, \mathbb{I}]&&=\int_\Sigma\,d^2 z\,\left[\;\boldlangle (g^{-1}\dd_z\epsilon\, g)_+,\,(J_-)_{\bar{z}}\boldrangle_{\mathbb{G}}+\boldlangle (J_+)_z, (g^{-1}\dd_{\bar{z}}\epsilon\, g)_-\boldrangle_{\mathbb{G}}\;\right]=\\
&&=\int_\Sigma\,d^2 z\,\left[\;\boldlangle g^{-1}\dd_z\epsilon\, g,\,(J_-)_{\bar{z}}\boldrangle_{\mathbb{G}}+\boldlangle (J_+)_z, g^{-1}\dd_{\bar{z}}\epsilon\, g\boldrangle_{\mathbb{G}}\;\right]
\eear
To pass to the second line we have used the following properties of the metric: $\boldlangle \mathfrak{h}, \mathfrak{m}_\pm\boldrangle_{\mathbb{G}}=0$, $\boldlangle \mathfrak{m}_\pm, \mathfrak{m}_\pm\boldrangle_{\mathbb{G}}=0$. The latter is a consequence of the Hermiticity of $\mathbb{G}$. Indeed, by definition of Hermiticity, $\boldlangle \mathbb{I}\circ u, \mathbb{I}\circ v\boldrangle_{\mathbb{G}}=\boldlangle u, v \boldrangle_{\mathbb{G}}$ for $u, v \in \mathfrak{m}$. Then, $\boldlangle \mathfrak{m}_+, \mathfrak{m}_+\boldrangle_{\mathbb{G}}=0$, since $\mathbb{I}\circ \mathfrak{m}_+=i\, \mathfrak{m}_+$. In other words, $\mathfrak{m}_+$ (and $\mathfrak{m}_-$) is an isotropic subspace of $\mathfrak{m}$. Using the invariance of the Killing metric $\mathbb{G}$ under the adjoint action of $G$, $\boldlangle gag^{-1}, gbg^{-1} \boldrangle_{\mathbb{G}}=\boldlangle a, b\boldrangle_{\mathbb{G}}$, we obtain
\bea
\delta\mathcal{S}[\mathbb{G}, \mathbb{I}]=\int_\Sigma\,d^2 z\,\left[\;\boldlangle\dd_z\epsilon,\,g (J_-)_{\bar{z}}g^{-1}\boldrangle_{\mathbb{G}}+\boldlangle \dd_{\bar{z}}\epsilon, g (J_+)_z g^{-1}\boldrangle_{\mathbb{G}}\;\right].
\eea
Using the non-degeneracy of $\mathbb{G}$, we deduce the conservation (on the e.o.m.) of the Noether current, defined as follows:
\bea\label{Scurr}
K=g\cdot\;\underset{:=S}{\underbrace{2\big((J_+)_z dz+(J_-)_{\bar{z}} d\bar{z}\big)}} \;\cdot g^{-1}\,=g S g^{-1}
\eea
Since the target space $\mathcal{M}=G/H$ is homogeneous, the equations of motion of the model are equivalent to its conservation:
\bea\label{curcons}
d\ast K=0.
\eea
In order to be able to build a family of flat connections we require that $K$ be flat (This will be used in (\ref{connection})-(\ref{flatnessAu}) below.):
\bea \label{curflat}
dK-K\wedge K=0.
\eea
We have to show, of course, that it is possible to satisfy this relation. Equations (\ref{curcons})-(\ref{curflat}) may be rewritten in terms of $S$ (introduced in (\ref{Scurr})) as follows:
\bear
d\ast S+\{J, \ast\, S\}=0\\
dS+\{J-{1\over 2} S, S\}=0
\eear
One checks directly that $J-{1\over 2} S=(J_-)_zdz+(J_+)_{\bar{z}}d\bar{z}+J_0$. Therefore
\bear \label{eqcons}
&&d\ast S+\{J, \ast \,S\}=\\ \nonumber&&=-2i dz\wedge d\bar{z}\;\;\big(\mathscr{D}_{\bar{z}} (J_+)_z-[(J_+)_z, (J_+)_{\bar{z}}]+ \mathscr{D}_z (J_-)_{\bar{z}}+[(J_-)_z, (J_-)_{\bar{z}}]\big)\\ \label{eqflat}
&&dS+\{J-{1\over 2} S, S\}=\\ \nonumber&&=-2dz\wedge d\bar{z}\;\; \big( \mathscr{D}_{\bar{z}} (J_+)_z-[(J_+)_z, (J_+)_{\bar{z}}]-\mathscr{D}_z (J_-)_{\bar{z}}-[(J_-)_z, (J_-)_{\bar{z}}]\big)
\eear
$\mathscr{D}$ is the covariant derivative for the gauge group $H$: $\mathscr{D}_jM_k:=\dd_jM_k+[(J_0)_j, M_k]$ ($j, k= z, \bar{z}$). We have used the definition of the Hodge star $\ast dz=i\,dz, \;\ast d\bar{z}=-i\,d\bar{z}$.

\vspace{0.3cm}
The conditions (\ref{eqcons})-(\ref{eqflat}) are equivalent, if
\bea\label{algcond}
[\mathfrak{m}_+, \mathfrak{m}_+]\subset \mathfrak{m}_+,\quad\quad [\mathfrak{m}_-, \mathfrak{m}_-]\subset \mathfrak{m}_-
\eea
(The commutators $[(J_+)_z, (J_+)_{\bar{z}}]$ and $[(J_-)_z, (J_-)_{\bar{z}}]$ then lie in the subspaces $\mathfrak{m}_+$, $\mathfrak{m}_-$ respectively.) In this case the equations take the form
\bear
\mathscr{D}_{\bar{z}} (J_+)_z-[(J_+)_z, (J_+)_{\bar{z}}]=0\\
\mathscr{D}_z (J_-)_{\bar{z}}+[(J_-)_z, (J_-)_{\bar{z}}]=0
\eear
If the metric $\mathbb{G}$ is Hermitian w.r.t. the chosen almost complex structure $\mathbb{I}$,
the conditions (\ref{algcond}) are equivalent to the integrability of $\mathbb{I}$. Indeed, integrability of the almost complex structure means, in general, that $[\mathfrak{m}_+, \mathfrak{m}_+]\subset \mathfrak{m}_+\oplus\, \mathfrak{h}$. To see this, note that the almost complex structure $\mathbb{I}$ has been defined by $\mathbb{I}\circ J_\pm=\pm i \, J_\pm$. Integrability of $\mathbb{I}$ is equivalent to the statement that (anti)-holomorphic forms generate a differential ideal, i.e. that $d(J_-)_a\sim \sum\limits_b \,R_{ab}\wedge (J_-)_b$ for some one-forms $R_{ab}$. Since $dJ+J\wedge J=0$, we have $$dJ_-=\big[-J_0\wedge J_0+(\textrm{terms involving}\,J_-) - J_+\wedge J_+\big]_{\mathfrak{m}_-}\,.$$ Therefore integrability of $\mathbb{I}$ requires that $[J_+\wedge J_+]_{\mathfrak{m_-}}=0$, i.e. $[\mathfrak{m}_+, \mathfrak{m}_+]\subset \mathfrak{m}_+\oplus \mathfrak{h}$. We see that the conditions (\ref{algcond}) define an integrable complex structure. Conversely, suppose we have an integrable complex structure on $G/H$, and $\mathfrak{m}_\pm$ are its holomorphic and anti-holomorphic subspaces. Then, $[a, b]=c+\gamma$, where $a, b, c\in \mathfrak{m}_+$ and $\gamma \in \mathfrak{h}$. Since $\boldlangle \mathfrak{m}_+, \mathfrak{h}\boldrangle_{\mathbb{G}}=0$, taking the scalar product with an arbitrary element $\gamma'\in \mathfrak{h}$, one obtains $\boldlangle \gamma', [a,b] \boldrangle_{\mathbb{G}}=\boldlangle \gamma', \gamma \boldrangle_{\mathbb{G}}$. Using $\boldlangle [a, \gamma'], b \boldrangle_{\mathbb{G}}+\boldlangle \gamma', [a, b] \boldrangle_{\mathbb{G}}=0$, we obtain $\boldlangle \gamma', \gamma \boldrangle_{\mathbb{G}}=-\boldlangle [a, \gamma'], b \boldrangle_{\mathbb{G}}=\boldlangle a', b \boldrangle_{\mathbb{G}}$, where $a'=[\gamma', a]\in \mathfrak{m}_+$. As discussed earlier, $\mathfrak{m}_+$ is isotropic if the metric $\mathbb{G}$ is Hermitian, therefore $\boldlangle \gamma', \gamma \boldrangle_{\mathbb{G}}=0$ for all $\gamma'\in \mathfrak{h}$, leading to $\gamma=0$ due to non-degeneracy of $\mathbb{G}$. We conclude that $[\mathfrak{m}_+, \mathfrak{m}_+]\subset \mathfrak{m}_+$.
\vspace{0.3cm}

Consider now the following family of connections $A_u$, indexed by a parameter $u\in \CC^\ast$:
\bea\label{connection}
A_u={1-u \over 2}\,K_z dz+{1-u^{-1}\over 2} \,K_{\bar{z}}d\bar{z}\;.
\eea

Conservation and flatness of the Noether current $K$, eqs. (\ref{curcons})-(\ref{curflat}), imply that $A_u$ is flat for all $u$:
\bea\label{flatnessAu}
dA_u-A_u\wedge A_u=0\quad\textrm{for all}\quad u\in \CC^\ast\;.
\eea

\subsection{Symmetric spaces}\label{symmspace}

As discussed in Section \ref{lagrsec}, Hermitian symmetric spaces satisfy the requirements~(\ref{prop}) and hence fall in our class. Sigma-models with such target spaces are known to be integrable, however the canonical Lax connection in this case is apparently different from the one in (\ref{connection}). Indeed, the connection usually employed in the analysis of $\sigma$-models with symmetric target spaces has the form
\bea
\tilde{A}_\lambda={1-\lambda \over 2}\,\tilde{K}_z dz+{1-\lambda^{-1}\over 2} \,\tilde{K}_{\bar{z}}d\bar{z},
\eea
where $\tilde{K}$ is the Noether current derived using the canonical action
\bea\label{actioncan}
\mathcal{S}[\mathbb{G}]=\int_\Sigma\,d^2 z\,\|\dd X\|^2_{\mathbb{G}}\;.
\eea
In the case of a Hermitian symmetric target space the difference between the two actions, (\ref{action}) and (\ref{actioncan}), is a topological term:
\bear
&&\mathcal{S}[\mathbb{G}, \mathbb{I}]-\mathcal{S}[\mathbb{G}]=\int_\Sigma\,X^\ast \omega,\\ \nonumber
&&\textrm{where} \quad d\omega=0\quad\textrm{if}\;\;\mathcal{M}\;\textrm{is symmetric.}
\eear
Therefore the two actions lead to the same equations of motion. Nevertheless, the Noether currents $K$ and $\tilde{K}$ are different. Indeed, let $\mathfrak{g}=\mathfrak{h}\oplus\mathfrak{m}$ with $\boldlangle \mathfrak{m},  \mathfrak{h}\boldrangle_{\mathbb{G}}=0$ be the standard decomposition of the Lie algebra $\mathfrak{g}$. The canonical Noether current $\tilde{K}$ is defined as follows:
\bea\label{noethercan}
\tilde{K}=2\,g\cdot\big[g^{-1}dg\big]_{\mathfrak{m}}\cdot g^{-1}
\eea
The notation $\big[\ldots\big]_\mathfrak{m}$ means that one should take the projection on the linear subspace $\mathfrak{m}\subset \mathfrak{g}$. This is certainly different from (\ref{Scurr}), since it follows from (\ref{noethercan}) that $\tilde{K}$ may be written as (taking into account that $\mathfrak{m}=\mathfrak{m}_+\oplus \mathfrak{m}_-$)
\bea
\tilde{K}=2\,g\cdot\left(J_++J_-\right)\cdot g^{-1}=2\,g\cdot \big(((J_+)_z+(J_-)_z)\, dz+((J_+)_{\bar{z}}+(J_-)_{\bar{z}})\,d\bar{z}\big)\cdot g^{-1}
\eea
On the other hand, $\tilde{K}$ is also flat: $d\tilde{K}-\tilde{K}\wedge \tilde{K}=0$. Checking this property does not, in fact, require using the equations of motion -- it is purely a consequence of the structure of the Lie algebra of the symmetric space (in particular, the fact that $[\mathfrak{m}, \mathfrak{m}]\subset \mathfrak{h}$). Moreover, the flatness condition may be solved, in this case, in a local fashion\footnote{`Local' means here that $\widehat{g}$ is a local function of the fields of the model.}:
\bea\label{Cartancurr}
\tilde{K}=-\widehat{g}^{-1}d\widehat{g},\quad \textrm{where}\quad \widehat{g}=\sigma(g) g^{-1},
\eea
$\sigma$ being Cartan's involution on the Lie group $G$. By definition, $\sigma$ is a group homomorphism, $\sigma(g_1 g_2)=\sigma(g_1)\sigma(g_2)$, and $\sigma(h)=h$ for $h\in H$. The formula $\widehat{g}=\sigma(g) g^{-1}$, viewed as a map $g\in G/H \to \widehat{g}\in G$, describes the Cartan embedding
\bea\label{Cartembed}
G/H \hookrightarrow G\;.
\eea

Flatness and conservation of the current $\tilde{K}$ lead to the flatness of the family $\tilde{A}_\lambda$. A question naturally arises of what the relation between $A_u$ and $\tilde{A}_\lambda$ is. The answer is that the connections $A_u$ and $\tilde{A}_\lambda$ are gauge-equivalent, if one makes the following identification of spectral parameters:
\bea\label{spectrpar}
\lambda=u^{1/2}\;.
\eea
It follows that, whenever $|u|=1$, one has $|\lambda|=1$, therefore if one connection is unitary, the other is as well. Relation (\ref{spectrpar}) is also consistent with the analysis in \cite{Hitchin} of the limiting behavior of the holonomies of the connection around the cycles of $\Sigma$ as $u\to 0$ (or ${u\to\infty}$). The gauge transformation $\Omega$ relating $A_u$ and $\tilde{A}_\lambda$, 
\bea
\tilde{A}_\lambda=\Omega A_u \Omega^{-1}-\Omega d\Omega^{-1},
\eea
may be constructed explicitly. The following formula holds for the case when the target-space is the Grassmannian $G_{n_1, n_1+n_2}:={SU(n_1+n_2)\over S(U(n_1)\times U(n_2))}$ and the complex structure is chosen so that it splits $\mathfrak{m}$ as $\mathfrak{m}=\left(\begin{smallmatrix}
 0 & \mathfrak{m}_+\\\mathfrak{m}_- & 0 \end{smallmatrix}\right)$:
\bea
\Omega=g\Lambda g^{-1}, \quad \textrm{where}\quad \Lambda=\lambda^{{1\over 2}{n_1-n_2\over  n_1+n_2}}\;\mathrm{diag}(\underset{n_1}{\underbrace{\lambda^{-1/2}, \ldots, \lambda^{-1/2}}}, \underset{n_2}{\underbrace{\lambda^{1/2}, \ldots, \lambda^{1/2}}})
\eea
In terms of the `dynamical' projectors $\Pi_{n_1}$, $\Pi_{n_2}$ on the subspaces of $\CC^{n_1+n_2}$ of the corresponding dimensions, $\Omega$ can also be written as
\bea
\Omega=\lambda^{{1\over 2}{n_1-n_2\over  n_1+n_2}}\;\left(\lambda^{-1/2} \,\Pi_{n_1}+\lambda^{1/2} \,\Pi_{n_2}\right)
\eea
In particular, at the point $\lambda=-1$ we obtain
\bea
\Omega|_{\lambda=-1}=e^{-i\pi {n_2\over n_1+n_2}}\left(\Pi_{n_1}-\Pi_{n_2}\right),
\eea
which is the explicit representation of Cartan's embedding (\ref{Cartembed}) for the case
\bea
G_{n_1, n_1+n_2}\hookrightarrow SU(n_1+n_2)\;.
\eea
At this point $u=1$. Since $A_{u=1}\equiv 0$, one has $\tilde{A}_{\lambda=-1}=\tilde{K}=-\Omega d\Omega^{-1}$, which coincides with representation (\ref{Cartancurr}), provided $\Omega=\widehat{g}^{-1}$.

\section{Discussion}

In this paper we showed that the $\sigma$-models with complex homogeneous target spaces, defined by action (\ref{action}), have the following property: their equations of motion may be rewritten as the flatness condition for a one-parameter family of connections. This representation is ubiquitously encountered in the realm of integrable models in two dimensions, and is often believed to be the characteristic, cornerstone property of these models \cite{Uhlenbeck, Devchand, Hitchin}. As we saw in Section~\ref{symmspace}, in the case of symmetric target spaces the proposed flat connection is gauge-equivalent to the conventional one, upon redefinition of the spectral parameter (\ref{spectrpar}).

\vspace{0.3cm}When the target-space is the flag manifold $\mathcal{M}=\frac{SU(3)}{S(U(1)^3)}$ and the worldsheet is a sphere, $\Sigma=\CP^1$, earlier \cite{Bykov2} we have been able to construct all solutions to the e.o.m. of the model (\ref{action}). The zero-curvature representation might allow to construct explicit solutions of the model (\ref{action}), in general.

\vspace{0.3cm}
Interestingly, the action (\ref{action}) coincides in form with the bosonic part of the action introduced in \cite{Witten}. In that paper the main focus was on a particular nilpotent supersymmetry of the full action, generated by a supercharge $\mathcal{Q}$, $\mathcal{Q}^2=0$. It was shown that the correlation functions of $\mathcal{Q}$-closed observables in those models could be computed. They are related to the moduli spaces of $\mathbb{I}$-holomorphic curves in the target space. (The holomorphic curves are absolute minima of the action (\ref{action}); the action vanishes on them.) Moreover, the correlation functions of $\mathcal{Q}$-closed observables are `topological', meaning that they do not change under smooth deformations of the metric $\mathbb{G}$ and of the complex structure $\mathbb{I}$. On the other hand, the zero-curvature representation, which we have constructed in the present paper, might allow to find other solutions of these models, not just the holomorphic curves. This zero-curvature representation only holds, however, if we choose the Killing metric and a compatible complex structure on the target space $\mathcal{M}$ of the $\sigma$-model.

\vspace{0.3cm}
\textbf{Acknowledgements.}
{\footnotesize
I would like to thank C.~Devchand and V.~Pestun for discussions.
I am indebted to Prof.~A.A.Slavnov and to my parents for support and encouragement. My work was supported in part by the grant RFBR 14-01-00695-a.
}

\vspace{0.7cm}
\begingroup
    \setlength{\bibsep}{4pt}
   \bibliography{nons}
\bibliographystyle{ieeetr}
\endgroup

\end{document}